\documentclass[final,3p,times,twocolumn]{elsarticle}
\usepackage{epsfig}

\journal{Physics Letters B}

\begin{document}

\begin{frontmatter}

\title{Search for Narrow Nucleon Resonance in $\gamma p\to \eta p$}

\author[gat,hel]{A.V.~Anisovich}

\author[hel]{E.~Klempt}

\author[kuz1,boh]{V.~Kuznetsov}

\author[gat,hel]{V.A.~Nikonov}

\author[gat,boh]{M.V.~Polyakov\corref{cor1}}
\ead{maxim.polyakov@tp2.rub.de}

\author[gat,hel]{A.V.~Sarantsev}
\ead{andsar@hiskp.uni-bonn.de}

\author[hel]{U.~Thoma}

\cortext[cor1]{Corresponding author}

\address[gat]{Petersburg Nuclear Physics Institute, Gatchina, St.Petersburg 188300, Russia}
\address[hel]{Helmholtz-Institut fu\"r Strahlen- u. Kernphysik, Universit\"at Bonn, Germany}
\address[kuz1]{Institute for Nuclear Research, 117312, Moscow, Russia}
\address[boh]{Institut f\"ur Theoretische Physik II, Ruhr-Universit\"at Bochum, D-44780 Bochum, Germany}

\begin{abstract}
Results of a partial wave analysis of new high-statistics data on
$\gamma p\to p\eta$ from MAMI are presented. A fit using known broad
resonances and only standard background amplitudes can not describe
the relatively narrow peaking structure in the cross section in the
mass region of 1660-1750~MeV which follows a minimum. An improved
description of the data can be reached by either assuming the
existence of a narrow resonance at a mass of about 1700~MeV with
small photo-coupling or by a threshold effect. In the latter case
the observed structure is explained by a strong (resonant or
non-resonant) $\gamma p\to\omega p$ coupling in the $S_{11}$ partial
wave. When the beam asymmetry data, published by part of the GRAAL
collaboration, are included in the fit, the solution with a narrow
$P_{11}$ state is slightly preferred. In that fit, mass and width of
the hypothetical resonance are determined to $M\sim$1694~MeV and
$\Gamma\sim 40$~MeV, respectively, and the photo-coupling to
$\sqrt{{\rm Br}_{\eta N}} A_{1/2}^p \sim 2.6\cdot
10^{-3}$~GeV$^{-1/2}$. High precision measurements of the target
asymmetry and $F$-observable are mandatory to establish the possible
existence of such a narrow state and to provide the necessary
information to define which partial wave is responsible for the
structure observed in the data.
\end{abstract}

\begin{keyword}


\end{keyword}

%\pacs{11.80.Et, 14.20.Gk, 14.20.Pt}
\end{frontmatter}

All nucleon resonances listed in the Review of Particle Properties
\cite{PDG} have rather large widths of $\Gamma\gtrsim 100$~MeV. Such
widths of nucleon resonances are natural for the constituent  quark
model picture of baryons (see the recent review of the baryon
spectroscopy and quark model ideas in Ref.~\cite{Klempt:2009pi}).
The chiral quark soliton model ($\chi$QSM) \cite{dia} challenges the
constituent quark model picture of nucleon resonances. Here, a SU(3)
anti-decuplet of light and narrow baryons is predicted. In
particular, the existence of the $P_{11}$ ($J^P=\frac 12^+$) nucleon
state, much narrower ($\le$40 MeV ) than normal nucleon excitations
of similar mass, was predicted~\cite{dia,dia1,arndt,michal}.

Ref.~\cite{max} predicts that a nucleon resonance from the
anti-decuplet is excited predominantly by photons off neutrons,
whereas its photo-excitation off protons is strongly suppressed.
Such a pattern, if observed, provides an imprint of the exotic
nature of a state. A partial wave analysis (PWA) of data on elastic
$\pi N$ scattering showed that the existing data on $\pi N$
scattering can tolerate a narrow $P_{11}$ resonance at a mass around
1680~MeV if its $\pi N$ partial decay width is below $0.5$~MeV
\cite{arndt}. Such a suppression of the $\pi N$ decay channel is
predicted in $\chi$QSM \cite{dia,arndt,michal}. The fact that the
excitation of the anti-decuplet nucleon in $\pi N$ and $\gamma p$
collisions is expected to be very weak makes the search of the
anti-decuplet nucleon a challenging task. Firstly, one needs high
precision and high resolution data\footnote{See \cite{epecur} where
the required precision and resolution of the $\pi N$ scattering data
to reveal the anti-decuplet nucleon is discussed.}. Secondly, a
detailed PWA of the data needs to be performed to reveal a weak
resonance signal. In contrast to $\gamma p$ reactions, the signal of
the anti-decuplet nucleon is expected to be rather sizable in
$\gamma n$ collisions \cite{max}.

Recently a peak structure has been observed in
$\eta$-photoproduction of the neutron, at $W\sim 1680$~MeV, by
GRAAL~\cite{gra0,gra1}, CBELSA/TAPS \cite{kru, Jaegle:2011sw},
LNS~\cite{kas}, and Crystal Ball/TAPS~\cite{wert}. Various
explanations for the structure have been proposed in the literature.
The structure can be produced by interference effects e.g. in the
$S_{11}$-wave as suggested in \cite{Anisovich:2008wd}. In
\cite{Shklyar:2006xw} the narrow structure was explained in terms of
coupled channel effects related to the $S_{11}(1650)$ and the
$P_{11}(1710)$-resonances. In the K-matrix coupled channel approach
of \cite{Shyam:2008fr} interference between $S_{11}(1535)$,
$S_{11}(1650)$, $P_{11}(1710)$, and $P_{13}(1710)$ leads to the
peak-like structure. A similar explanation but with an additional
contribution from a rather narrow $D_{13}(1700)$ state was suggested
in \cite{Zhong:2011ti}. In \cite{Doring:2009qr} the effect was
explained by interference of the two $S_{11}$ states and a strong
cusp effect at the $K\Lambda$ and $K\Sigma$ thresholds.

A very different interpretation is provided in \cite{az,tia}. It was shown
that in $\eta$-photoproduction of the neutron  the signal can be described by
the contribution of a narrow resonance. In $\gamma n$ collisions
(with non-suppressed exit channels such as $\eta n$, $\gamma n$,
$K_S\Lambda$, etc.) the anti-decuplet state should be seen as a
clear narrow peak in the cross section \cite{max}.

Recently \cite{Compton}, quasi-free Compton scattering on the
neutron in the energy range of $E_{\gamma}=750 - 1500$~MeV was
studied. The data indicate the existence of a narrow ($\Gamma\sim
35$~MeV) peak at $W\sim 1685$~MeV. The peak is absent in the Compton
scattering off protons as well as in the reactions $\gamma n\to
\pi^0 n$ and $\gamma p\to \pi^0 p$. The latter observation implies
that the putative narrow resonance should have a very small $\pi N$
partial width, in agreement with the modified PWA of
Ref.~\cite{arndt} and with theoretical expectations for the
anti-decuplet nucleon \cite{dia,dia1,arndt,michal}.

If the peak at 1680\,MeV would be due to narrow nucleon resonance,
it should as well contribute to the $\gamma p$ channel. However, if
it is related to the antidecuplet state, its contribution is
predicted to be considerably suppressed. Possibly, it can
nevertheless be seen in observables exploiting its interference with
a strong smoothly varying background. The corresponding signal would
then not necessarily look like a peak but may appear rather as a
structure oscillating with energy, or as a dip.

A first search of the putative anti-decuplet nucleon in $\gamma p\to
\eta p$ process was performed in Refs.~\cite{acta,jetp}. It was
found that the beam asymmetry $\Sigma$, published by part of the
GRAAL collaboration, exhibits  a structure around $W\sim1685$~MeV.
That structure looks like a peak at forward angles  which develops
more into an oscillating structure at larger scattering angles. Such
a behavior may occur by interference of a narrow resonance with a
smooth background. The observed structure was identified in
Refs.~\cite{acta,jetp} with the contribution of a resonance with
mass $M\sim 1685$~MeV, narrow width of $\Gamma\le 25$~MeV, and small
photo-coupling of $\sqrt{{\rm Br}_{\eta N}} A_{1/2}^p \sim
(1-2)\cdot 10^{-3}$~GeV$^{-1/2}$.

Recently, the Crystal Ball Collaboration at MAMI published high
precision data on $\eta$ photoproduction on the free proton
\cite{Mainz}. The cross section was measured in fine steps in photon
energy. The measured cross section exhibits a minimum at masses
around $\approx$ 1680~MeV followed by a slight maximum around
1700~MeV. The best fit to the data was achieved with a new version
of SAID (GE09) \cite{Mainz}. The authors interpret the fit as
evidence against the existence of a narrow bump at $1680$~MeV in
this reaction. However, inspection of their fit reveals a systematic
excess of data above their fit curves in the $1710-1730$~MeV region.

In \cite{KPT} the data of Ref.~\cite{Mainz} were interpreted as
indication for a nucleon resonance with mass of $M\sim 1685$~MeV, a
narrow width of $\Gamma\leq 50 $~MeV, and a small resonance
photo-coupling  in the range of $\sqrt{{\rm Br}_{\eta N}} A_{1/2}^p
\sim (0.3-3)\cdot 10^{-3}$~GeV$^{-1/2}$. In this case no PWA of the
data was done as needed to decide whether or not a resonance occurs
in a certain partial wave.

In this Letter we report on a PWA of the new MAMI
data \cite{Mainz} which aims to trace the physical origin of the
small deviation between data and the SAID (GE09) fit.
The MAMI data are incorporated into the large data base on pion- and
photo-induced reactions which is exploited in the Bonn-Gatchina
coupled-channel analysis. PWA methods are described in detail in
\cite{Anisovich:2004zz,Klempt:2006sa,Anisovich:2006bc,Anisovich:2007zz,Anisovich:2005tf},
a list of data used in the most recent analysis is given in
\cite{BG}. In the latter work, two main solutions, BG2010-01 and
BG2010-02, were found. These solutions have very close parameters
for the resonances below 1800 MeV but differ in couplings and pole
positions for higher-mass states. The description of the
observables, multipoles and $\pi N$ transition amplitudes with these
solutions can be downloaded from \cite{BG_www}.

\begin{table*}
\caption{\label{D=6_I=0}The description of the MAMI \cite{Mainz} and
beam asymmetry data \cite{acta,jetp}. For the $P_{11}$ narrow state
two solutions with positive $P_{11}(+)$ and negative $P_{11}(-)$
interferences are given. The $\chi^2_{\rm sel}/N_{\rm dat}$
corresponds to the $\chi^2$ per point for the mass region 1660-1750
MeV. The $\chi^2_{\Sigma}/N_{\rm dat}$ corresponds to the $\chi^2$
per point for the beam asymmetry data from \cite{acta,jetp} and from
\cite{Bartalini:2007fg}. Masses and widths are in MeV units. The
photocouplings are in units of $10^{-3}$~GeV$^{-1/2}$.}
\begin{center}
\begin{tabular}{|c|c|c|c|c|c|c|c|c|}
\hline
Resonance  & Mass & $\Gamma_{\rm tot}$ & $\sqrt{{\rm
Br}_{\eta N}} A_{1/2}^p$ & $\sqrt{{\rm Br}_{\eta N}} A_{3/2}^p$ &
$\chi^2_{\rm tot}/N_{\rm dat}$ & $\chi^2_{\rm sel}/N_{\rm dat}$
&$\chi^2_{\Sigma}/N_{\rm dat}$\cite{acta,jetp} & $\chi^2_{\Sigma}/N_{\rm dat}$
\cite{Bartalini:2007fg}\\
\hline
no res. &-&-&-&-&1.21&1.46& 1.52 & 1.40\\
\hline
$P_{11}(+)$ &1719&44&2.75&-&1.07&0.90 &1.44 & 1.49\\
\hline
$P_{11}(-)$ &1694&41&2.6 &-&1.13& 0.92 & 1.18&1.41\\
\hline
$P_{13}$&1728&72&2.45&4.5&1.02&0.93 & 1.47&1.37\\
\hline
$S_{11}$ &1696&31&0.77&-&1.11&1.10 & 1.42&1.52\\
\hline
$S_{11}(\omega p)$ &-&-&-&-&1.12&0.94 &1.40&1.52\\
\hline
\end{tabular}
\end{center}
\end{table*}
Both solutions show small but systematic deviations from the data of
Ref.~\cite{Mainz}, from the threshold up to 1750\,MeV. At higher
masses, the new data seem to be compatible with the Crystal Barrel
measurements \cite{Crede:2009zzb}. However, the latter data are
presented in larger energy bins. Using this data, we re-optimized
the parameters of the fit-solutions. In the present fit we included
MAMI data with effective errors by adding quadratically statistic
and systematic errors.

\begin{figure}[ht]
\centerline{
\epsfig{file=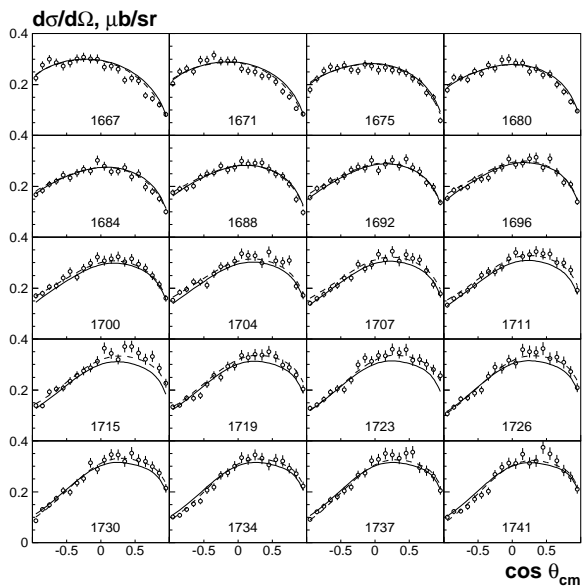,width=0.47\textwidth,clip=on}}
\caption{\label{eta_01_diff}Differential cross section $\gamma p\to
\eta p$ in the mass region 1660-1750 MeV \cite{Mainz} taken from the
Durham data base. The full curves correspond to the solution
BG2010-02M and the dashed curves to the solution $P_{11}(+)$. Errors
for experimental points are taken as square root from sum of
statistic and systematic errors.}
\end{figure}

\begin{figure}[ht]
\centerline{
\epsfig{file=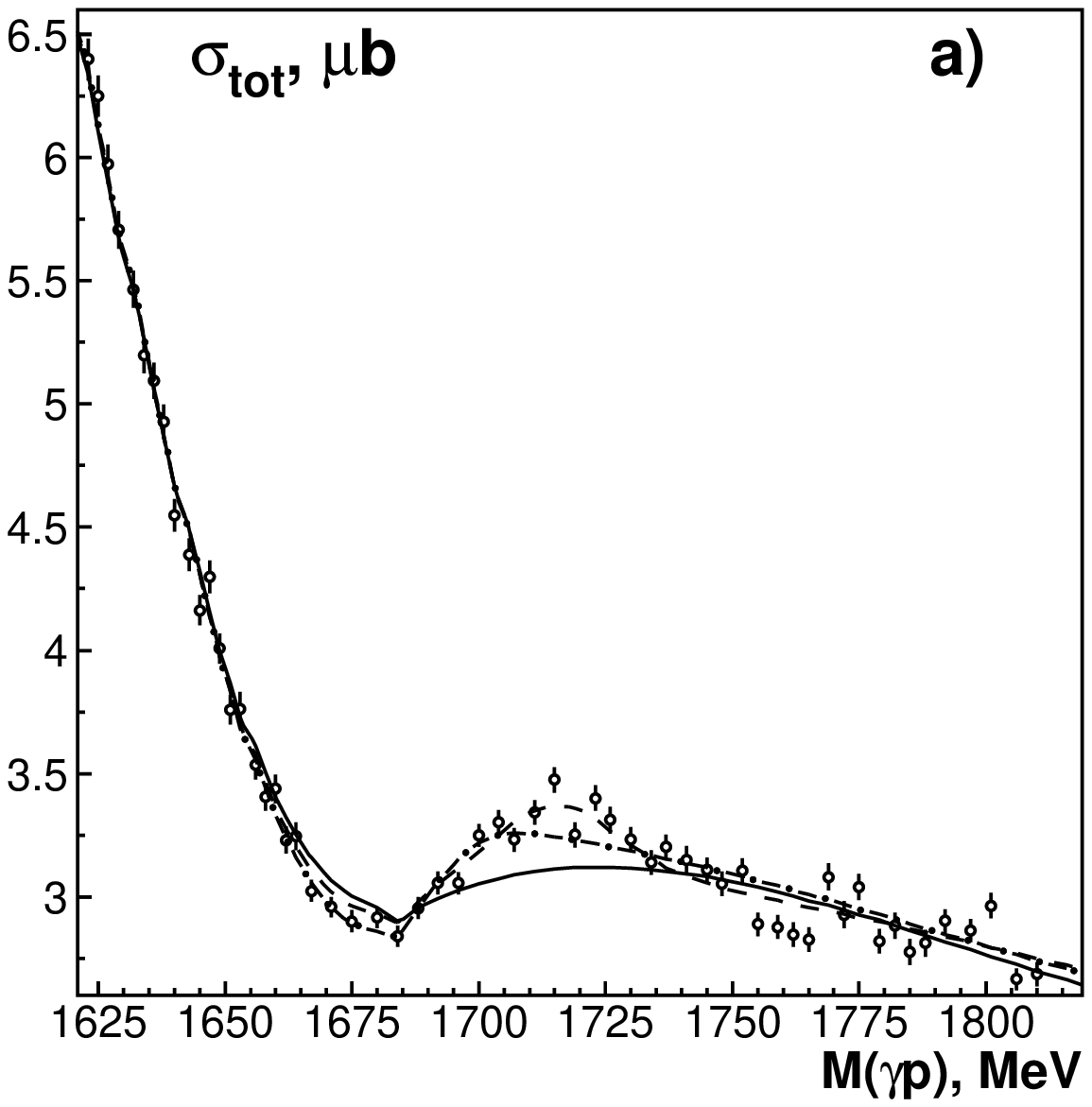,width=0.23\textwidth,clip=on}
\epsfig{file=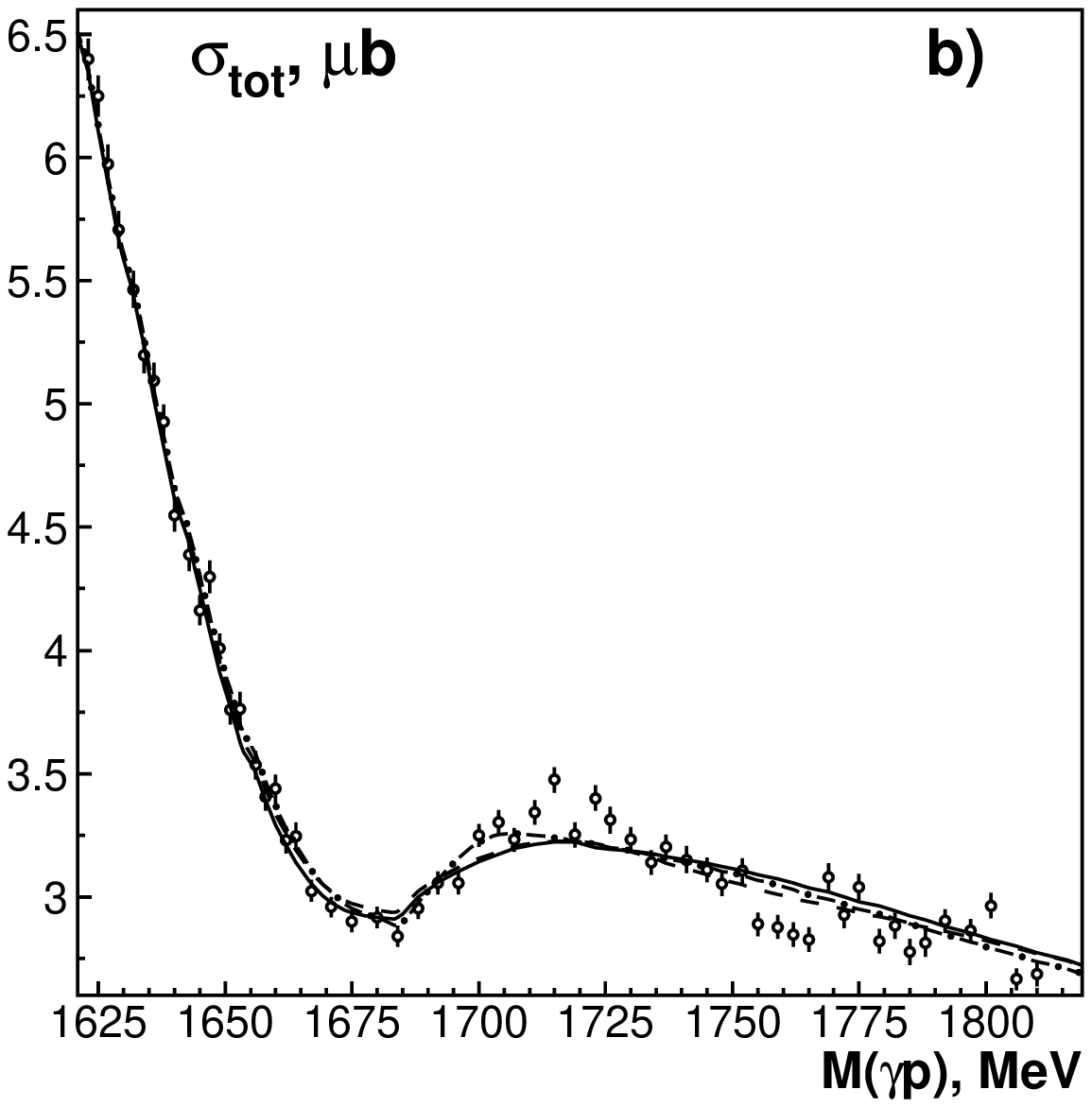,width=0.23\textwidth,clip=on} }
\caption{\label{eta_tot_p11} Comparison of the total cross section
$\gamma p\to \eta p$ calculated from the data \cite{Mainz} with
found solutions. a) The full curve corresponds to the solution
BG2010-02M, the dashed curve to the solution $P_{11}(+)$ and
dashed-dotted curve to the solution $P_{11}(-)$. b) The full curve
corresponds to the solution BG2010-02M with $\omega p$ channel
included, the dashed curve to the solution with the $P_{13}$ narrow
state and dashed-dotted curve to the solution with the $S_{11}$
narrow state.}
\end{figure}

Only a small redefinition of parameters was required to obtain a
rather good overall description of the MAMI data for both solutions
(BG2010-01 and BG2010-02) with a $\chi^2$ per data point of
$\chi^2/N_{\rm dat}=1.21$. However, in the $1660-1750$~MeV mass
interval and, especially, above 1700~MeV, the solutions still have
statistically significant deviations between data and fit. In this
interval, $\chi^2/N_{\rm dat}=1.46$ is considerably larger than in
other energy intervals. This is illustrated in
Fig.~\ref{eta_01_diff} for the differential cross sections and in
Fig~\ref{eta_tot_p11} for the total cross section. The differential
cross section is not described well in the angular region
$0<\cos\Theta<0.5$ for the energy bins at 1667, 1671 and 1700 to
1730 MeV.

The systematic deviations between data and fit show that our
solutions miss some physics in this region. The missing contribution
is likely either in $S$ or $P$ wave. A contribution either from the
$D$ or $F$ partial waves provides a complicated angular behavior,
which is not compatible with the smooth observed angular
distributions.

The most striking explanation of this phenomenon is the existence of
a narrow state with the mass around 1700 MeV. Indeed, including a
narrow state significantly improves the description of the $\gamma
p\to \eta p$ data in this mass region. Starting from the solution
BG2010-02M and assuming the contribution of a narrow state, we have
found four solutions which provided a similar $\chi^2$: two
solutions with a $P_{11}$ narrow state, one solution with a narrow
state in the $P_{13}$ partial wave and one solution with a narrow
$S_{11}$ resonance. In Table~\ref{D=6_I=0} our solutions are listed;
for each solution we give the total $\chi^2$, the $\chi^2_{\rm sel}$
for the data from Ref.~\cite{Mainz} in the energy interval of
$1660-1750$~MeV, and the $\chi^2_\Sigma$ for the data on beam spin
asymmetry from Refs.~\cite{acta,jetp} and from
Ref.~\cite{Bartalini:2007fg}.

The two solutions with a narrow $P_{11}$ resonance differ by the
interference of this state with other partial waves. For positive
interference (solution $P_{11}(+)$), the mass of the resonance
optimizes at 1719\,MeV, and the width at 44\,MeV. The solution
exhibits a clear peak at 1719\,MeV where the experimental total
cross section reaches its maximum. The solution with negative
interference (solution $P_{11}(-)$) describes better the region
around 1690\,MeV where the total cross section reaches the minimum.
In this solution the mass optimizes for 1694, the width for 41\,MeV
(see Table~\ref{D=6_I=0}). The differential cross section, the
solution BG2010-02M  without a resonance, and the solution with
$P_{11}(+)$ resonance are shown in Fig.~\ref{eta_01_diff}; the total
cross section in this mass region calculated from experimental data
points is compared with solution BG2010-02M and both $P_{11}$
solutions in Fig.~\ref{eta_tot_p11}a. Although the solution with
positive interference reproduces better the total cross section, the
$\chi_{\rm sel}^2$ for the fit to the differential cross section is
practically the same.

Indeed, the $P_{11}(+)$ solution solves the problem in the
description of the differential cross section in the forward angular
region, but slightly exceeds the differential cross section in the
backward region (see mass bins 1707,1711 and 1715 MeV). The obtained
values of the photo-couplings (see Table~I) in both solutions are in
a good agreement with estimates from Refs.~\cite{acta,jetp,KPT} and
are about 5 times smaller than the corresponding coupling for the
neutron obtained in Ref.~\cite{az} from the analysis of the GRAAL
data \cite{gra0,gra1}.

In the case of a $P_{13}$ narrow state, a notable improvement in the
description was obtained from a fit with $A_{3/2}^p\sim 2A_{1/2}^p$
and destructive interference with the other parts of the $P_{13}$
wave. The total cross section for this solution is shown in
Fig.~\ref{eta_tot_p11}. The mass of the $P_{13}$ state optimized at
1728 MeV and width at 72 MeV. This state is relatively broader than
the narrow states in other solutions and produces a rather
complicated interference with the rest of the $P_{13}$ wave. The
structure, if confirmed, might be related to the anomaly in the
$P_{13}$ wave reported by CLAS in electro-production of $\pi^+\pi^-$
pairs \cite{Ripani:2002ss}.

A narrow $S_{11}$ state only slightly improves the description of
the data in the 1660-1750\,MeV region. Only one solution with
destructive interference with the remaining $S_{11}$ partial wave
was found. The mass of this state optimized at $M=1685$\,MeV and
$\Gamma=30$\,MeV. The comparison of the experimental total cross
section and the result of the fit is shown in
Fig.~\ref{eta_tot_p11}b. The value $\sqrt{{\rm Br}_{\eta N}}
A_{1/2}^p$ for this solution is about 4 times smaller than that for
other solutions. There is no surprise here: this resonance
interferes with the largest partial wave and a small coupling can
produce a notable effect.

Another - more conventional - possibility to reproduce the narrow
structure is a contribution from the $\omega p$ threshold. It is
known that diffractive $\omega$ production is important; possibly,
its effect (due to pion exchange) is experienced already in the
threshold region. The $\omega p$ channel opens at 1720 MeV. Sizable
effects can be expected when $\omega$ and proton are in relative $S$
wave. This translates into $S_{11}$ and $D_{13}$ partial waves. It
is interesting to investigate whether the opening of this channel
can explain the structure observed in \cite{Mainz}. To check this
assumption we introduced the $\omega p$ channel in the K-matrix
parameterization of the $S_{11}$ partial wave and fitted the
couplings of the K-matrix poles and for non-resonant transition
$\gamma p\to \omega p$ as free parameters. The 650-1750\,MeV mass
region of the MAMI data is now described better than in the fit with
a narrow $S_{11}$ resonance. The $\chi^2$s for this fit are given in
Table~\ref{D=6_I=0}. The description of the beam asymmetry data
\cite{acta,jetp} and \cite{Bartalini:2007fg} practically can not be
distinguished from that obtained with the BG2010-02M solution and is
not shown. The inclusion of the $\omega p$ channel in the $D_{13}$
wave did not improve the description of the data.

The optimum strength for the non-resonant transition term $\gamma
p\to \omega p$ in the $S_{11}$ wave was found at a rather large
value. However, there is a large correlation between the parameter
representing the non-resonant transition amplitude and the
parameters representing decays of the two $N(1535)S_{11}$ and
$N(1650)S_{11}$ resonances into $p\omega$, and the resonant and
non-resonant contributions cannot be distinguished on the basis of
the data used here. The $\omega p$ parameters can likely be fixed
when data on $\gamma p\to \omega p$ are included in the data base.
This option is not yet included in our fitting program. We note that
the effects of the $K\Lambda$ and $K\Sigma$ thresholds are included
in our analysis automatically: the $KY$ photo and pion induced data
are an important part of our couple channel analysis. In our present
solutions we do not observe a significant effects from the $KY$
thresholds on the $eta$ photoproduction data.

\begin{figure}[ht]
\centerline{ \epsfig{file=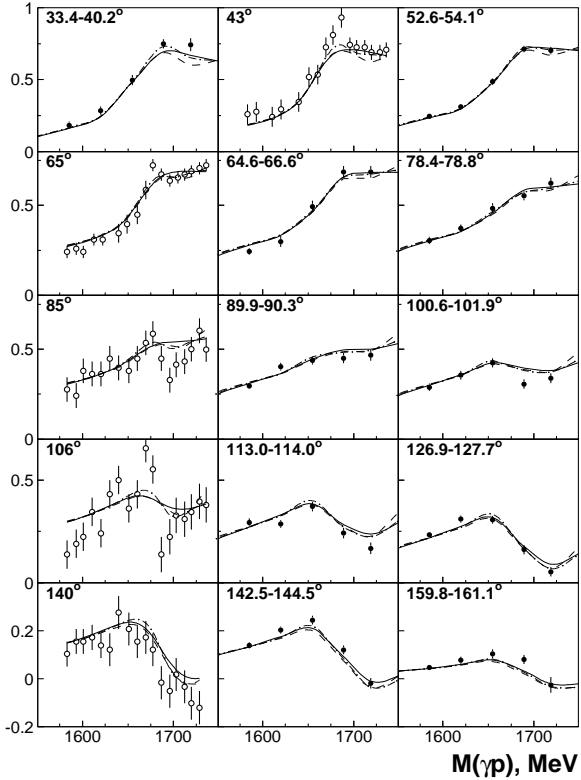,width=0.48\textwidth,clip=on}}
\caption{\label{eta_kuzn_p11} The description of the beam asymmetry
data (shown at fixed angles) with our solutions. The open circles
represent the data from \cite{acta,jetp} and full circles the data
from \cite{Bartalini:2007fg}. The center values of angular bins for
\cite{Bartalini:2007fg} depend on the energy and are given as
intervals (from the lowest energy to highest one). The full curve
corresponds to the solution BG2010-02M, dashed curve to the
$P_{11}(+)$ solution and dashed-dotted curve to the $P_{11}(-)$
solution.}
\end{figure}

Polarization observables are very useful to decide to which partial
wave the structure in MAMI data can be related. Polarization
measurements - including the GRAAL data on beam asymmetry for $\eta$
photoproduction~\cite{Bartalini:2007fg} are, of course, part of our
data base. Yet, the beam asymmetry results of~\cite{acta,jetp} -
from an alternative analysis of a limited sample of the same data
and published by part of the GRAAL collaboration only - was not
included. However, as mentioned above, additional evidence for a
narrow state with mass around 1685~MeV was reported from the
analysis of this data \cite{acta,jetp} and therefore we included
this data set with a relatively small weight in the present data
base. The description of the MAMI data appeared to be hardly
sensitive to inclusion of these data. Only the solution $P_{11}(-)$
produced a better description of the beam asymmetry data
\cite{acta,jetp} and even a slightly better ($\delta\chi=0.03$)
description of the MAMI data in the mass region 1650-1750 MeV. The
$\chi^2$ for the description of the MAMI data in the total mass
region, in the region of 1660-1750~MeV and for the beam asymmetry
data are given in Table~\ref{D=6_I=0}. The description of the beam
asymmetry data for three solutions (BG2010-02M, $P_{11}(+)$ and
$P_{11}(-)$) is shown in Fig.~\ref{eta_kuzn_p11}. The solutions with
the $P_{13}$  and $S_{11}$ narrow states shows only very small
deviations from the solution BG2010-02M.

The beam asymmetry data can hardly can distinguish between the
different solutions. At present, the beam asymmetry from
\cite{acta,jetp} might slightly favor the solution $P_{11}(-)$ but
the statistical significance does not enforce one of the solutions.
Future precise measurements of polarization observables are needed
to decide which of the solutions proposed here corresponds best to
reality.

\begin{figure}[ht]
\centerline{ \epsfig{file=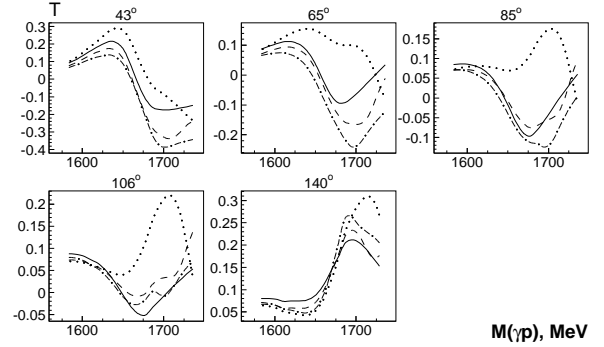,width=0.47\textwidth,clip=on}}
\caption{\label{eta_t} Prediction of the target asymmetry for $\eta$
photoproduction.  The full curves correspond to the solution with
$\omega p$ channel included to the $S_{11}$ partial wave, dashed
curves to the $P_{11}(+)$ solution, dashed-dotted curves to the
$P_{11}(-)$ solution and dotted curves to the $P_{13}$ solution.}
\end{figure}

\begin{figure}[ht]
\centerline{ \epsfig{file=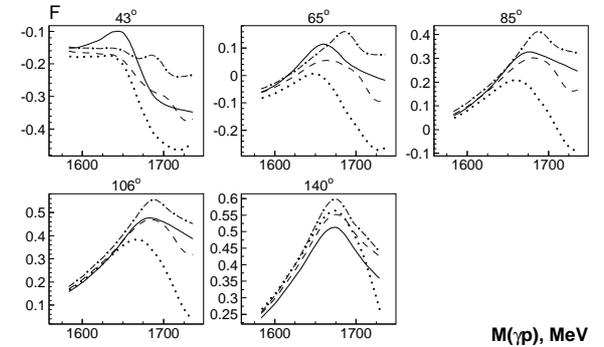,width=0.47\textwidth,clip=on}}
\caption{\label{eta_f} Prediction for the F-observable in the $\eta$
photoproduction.  The full curves correspond to the solution with
$\omega p$ channel included to the $S_{11}$ partial wave, dashed
curves to the $P_{11}(+)$ solution, dashed-dotted curves to the
$P_{11}(-)$ solution and dotted curves to the $P_{13}$ solution.}
\end{figure}

One of the most sensitive observables which would be able to distinguish
between the four solutions is the target asymmetry. The prediction
of this observable from our four solutions (with $\omega p$ channel
introduced in the $S_{11}$ wave, with the narrow $P_{11}$ states and
with the narrow $P_{13}$ state) is shown in Fig.~\ref{eta_t}.
Another interesting polarization observable sensitive to these
different solutions can be extracted from an experiment with a
transversely polarized target and circularly polarized beam. The
prediction of the so-called $F$-observable from the four solutions
discussed above is shown in Fig.~\ref{eta_f}. With such data, even
solutions with negative and positive interference between a narrow
$P_{11}$ resonance and the remaining wave could be separated if the
narrow state would indeed contribute to the data.

We conclude that the new high precision data on $\gamma p\to \eta p$
cross section of Ref.~\cite{Mainz} reveal an interesting structure
in the mass region of 1660-1750~MeV. The relatively smooth angular
distributions suggest that this structure can be interpreted within
the $P$ or $S$ waves. The threshold of the $\omega p$ channel may
effect the data and my contribute by a coupling of the two $S_{11}$
resonances to $\omega p$ and by a non-resonant $\gamma p\to p\omega$
transition strength. A good fit of the data is achieved when the
$\omega p$ channel is included even though the fit is unable to
decide which of the two mechanisms is more important. Assigning the
effect to the $P$-wave, the data can be explained only with
introduction of a narrow resonance, in particular when the data
\cite{acta,jetp} on the beam asymmetry $\Sigma$ are included. A
narrow $P_{11}$ resonance - interfering destructively within the
$P_{11}$ wave - would be preferred in this case.

High statistic polarization data on target asymmetry and on the
double polarization variable $F$ should provide the necessary
constraints to define which partial wave is responsible for the
structure observed in mass region of 1660-1750~MeV in the $p\eta$ cross
section. In the end it may provide the information needed to decide
whether or not a narrow baryon resonance exists in this mass range.

 {\it \small  The work has been supported by the DFG within SFB/TR16.
 M.V.P. is thankful to N. Sverdlova for
 comments on the text and to I. Strakovsky for correspondence.}

\end{document}